\documentclass[cha,
aip,
 amsmath,amssymb,
 reprint,%
]{revtex4-1}

\usepackage{graphicx}% Include figure files
\usepackage{dcolumn}% Align table columns on decimal point
\usepackage{bm}% bold math
\usepackage[utf8]{inputenc}
\usepackage[T1]{fontenc}
\usepackage{mathptmx}
\usepackage{etoolbox}

\makeatletter
\def\@email#1#2{%
 \endgroup
 \patchcmd{\titleblock@produce}
  {\frontmatter@RRAPformat}
  {\frontmatter@RRAPformat{\produce@RRAP{*#1\href{mailto:#2}{#2}}}\frontmatter@RRAPformat}
  {}{}
}%
\makeatother
\begin{document}

\preprint{AIP/123-QED}

\title[Stationary and switching synchronization states in an ensemble of four nonidentical phase oscillators with repulsive couplings]{Stationary and switching synchronization states in an ensemble of four nonidentical phase oscillators with repulsive couplings}

\author{A.V. Bobrova$^*$}
\email{alla.bobrova@unn.ru}
\author{G.V. Osipov}
\affiliation{
Department of Control Theory, Research and Education Mathematical Center "Mathematics for Future Technologies", Lobachevsky State University of Nizhny Novgorod, 23 Gagarin Avenue, Nizhny Novgorod, 603022, Russia
}

\date{\today}% It is always \today, today,
             %  but any date may be explicitly specified

\begin{abstract}
This study investigates the collective dynamics (phase synchronization, instantaneous frequencies synchronization and mean frequencies synchronization) in an ensemble of four nonidentical phase oscillators with repulsive coupling. We use the Kuramoto-Sakaguchi system of ordinary differential equations as our mathematical model. Depending on the coupling strength in the presence of a small mismatch of the natural frequencies, all possible states of frequency synchronization were found: 4:0 (global), 3:1, 2:2, 2:1:1 (cluster). It is shown that these states can be classified into two main types depending on the evolution of the instantaneous frequencies: stationary (4:0 and 2:2), characterized by constancy of phase mismatch and instantaneous frequencies, and switching (3:1 and 2:1:1), in which metastable processes with periodic switching of synchronous states are observed: for different time intervals, different types of locking of instantaneous frequencies of oscillator pairs and different types of phase mismatch were observed. A novel effect of collective dynamics -- self-organized emergence of slow-fast behavior in the ensemble of elements which, if decoupled, exhibit uniform temporal evolution was demonstrated. The rather long time intervals during which the distances between some oscillators stay nearly constant alternate with short epochs of rapid evolution, involving the break up of clusters and their subsequent reassembly. For the 4:0 and 2:2 states, analytical expressions for the synchronization frequencies were derived. The presence of bistability has been revealed depending on the initial conditions of different synchronization states at the same parameters: sets of individual frequencies and value of coupling strength.
\end{abstract}

\maketitle

\begin{quotation}
The collective dynamics of coupled phase oscillator ensembles exhibits a wide range of phenomena, including global and partial (clustered) synchronization. Although partial synchronization in such ensembles is a well-known phenomenon, the underlying mechanisms and the classification of states in systems with \textit{nonidentical} elements and \textit{repulsive} coupling require further investigation. In this paper we study a system of four phase oscillators and demonstrate that repulsive coupling combined with a small frequency mismatch leads not only to conventional stationary states (global 4:0 synchronization and pairwise cluster 2:2), but also to the emergence of switching states (clustered synchronization states 3:1 and 2:1:1) characterized by metastability and periodic switching of synchronous states.
\end{quotation}

\section{Introduction}

Studying synchronization states in ensembles of interacting oscillators across diverse systems is one of main area in nonlinear dynamics. These systems are encountered in biology~\cite{Nieto2019, Merritt2018,Prindle2011}, physics~\cite{Qian2008,Clerkin2014}, chemistry~\cite{Smelov2019}, neuroscience~\cite{Izhikevich2007,Goriely2020,Ashwin2016}, and other fields. Despite the wide area of practical applications, most of these systems can be reduced to systems of coupled phase oscillators~\cite{Pikovsky2001,Winfree1967,Kuramoto1984}. The primary factors complicating the analysis of phase system behavior are individual element dynamics, coupling strength, and interaction topology -- global, local, or nonlocal~\cite{Kuramoto2002,Soomin2025}.

An important tool for analyzing the behavior of coupled phase oscillators is the Kuramoto model~\cite{Kuramoto1975}. The classical model considers sinusoidal coupling proportional to $\sin{(\varphi_k - \varphi_j)}$, representing the simplest oscillator interaction. However, accounting for additional physical factors like coupling delays requires generalizing the classical model. The Kuramoto-Sakaguchi model~\cite{Sakaguchi1986} partly fulfills this requirement. Adding a parameter of phase shift $\alpha$ ($\sin{(\varphi_k - \varphi_j - \alpha)}$) enables modeling repulsive coupling and more complex interactions~\cite{Leon2019}. It should be stated that the distinction between attractive and repulsive coupling concerns oscillators with sufficiently close values of their respective phases. If action of the coupling drives the oscillators closer, the coupling is called attracting; if, in contrast, the originally close oscillators are driven apart, the coupling is repulsive.   

While the case of attractive coupling has been relatively well studied, the dynamics of systems with repulsive coupling demonstrates significantly richer and often counterintuitive behavior. The work by Tsimring et al.~\cite{Tsimring2005} showed that globally coupled identical oscillators with repulsive coupling evolve into a state with a zero mean field, while for non-identical oscillators, complete synchronization for $N>3$ turns out to be impossible. It is important to note that for identical oscillators with strictly sinusoidal coupling, the dynamics is partially integrable (Watanabe-Strogatz theory), which imposes strict limitations -- for example, it prohibits the formation of stable cluster states, and their observation in numerical experiments is often a discretization artifact~\cite{Gong2019}. Genuinely complex dynamics, including multiple attractors and complex synchronization states, emerges when the conditions of this integrability are violated. This violation can be achieved through different pathways: for instance, through a transition to excitable dynamics of the elements, where global repulsive coupling generates a multitude of periodic states via a transcritical heteroclinic bifurcation~\cite{Zaks2016}, or through the introduction of non-identity and mixed interactions, which leads to phenomena such as solitary states and quasiperiodic partial synchrony~\cite{Teichmann2019}. Several interesting result concerning dynamics, including chaos, in an ensemble of phase oscillators are presented in Refs.\cite{Popovych2005a,Maistrenko2005,Karatetskaia2025}

The present work builds, in part, upon the research results of Professor V.S. Afraimovich in the field of synchronization of oscillations in dynamical systems of various natures. He obtained results that have become classical in bifurcation theory, regular and chaotic synchronization, and structure formation. We also note the contribution of V.S. Afraimovich, whose research demonstrated that the stability of the structure of coupled oscillator networks depends on the network topology and the distribution of oscillator frequencies. His work provided a deeper understanding of the conditions under which collective phenomena, such as synchronization and chaos, emerge in systems of coupled oscillators, particularly in cases with complex topology and frequency heterogeneity~\cite{Afraimovich2004}.

The novelty of our work lies in the systematic classification of all five frequency and phase synchronization states for an ensemble of four nonidentical phase oscillators with repulsive coupling, in the identification of two qualitatively different types of synchronous dynamics (stationary and switching).

\section{Model}

The collective dynamics of an ensemble of coupled phase oscillators is described by the Kuramoto-Sakaguchi system~\cite{Sakaguchi1986}:

\begin{equation}\label{eq2.1}
    \dot{\varphi_j}=\gamma_j+\displaystyle\frac{d}{N}\sum^N_{k\neq j}{\sin{(\varphi_k-\varphi_j-\alpha)}},
\end{equation}
where $j=1,\dots,N$, $N$ is the number of oscillators, $\gamma_j$ are the natural frequencies of the oscillators, $d\in[0,30]$ is a parameter determining the coupling strength, and $\alpha$ characterizes the phase shift in the couplings. The interval $\alpha\in\Big(0,\displaystyle\frac{\pi}{2}\Big]$ corresponds to attractive coupling, whereas $\alpha\in\Big(\displaystyle\frac{\pi}{2},\pi\Big]$ corresponds to repulsive coupling.

We examine the system with repulsive coupling ($\alpha=1.6$), $N=4$, and nonidentical natural frequencies uniformly distributed over the interval $\gamma_j\in[\gamma-\Delta\gamma,\gamma+\Delta\gamma]$, where $\gamma=3,\Delta\gamma=0.01$. It should be noted that the value 3 can be replaced by any other real number, including zero.

\section{Synchronization}

In order to study the collective behavior of oscillators in the Kuramoto-Sakaguchi system, we will focus on various types of both phase and frequency synchronization.

We will distinguish the in-phase synchronization state, where
\begin{equation}\label{eq2}
    |\varphi_k (t)-\varphi_j (t)|<c
\end{equation}
and the anti-phase synchronization state, where
\begin{equation}\label{eq3}
    |\varphi_k (t)-\varphi_j(t)-\pi|<c.
\end{equation}

In the present work, to confirm the presence of synchronization of both types, inequalities \eqref{eq2} and \eqref{eq3} were satisfied for $c=0.001$ over a time interval of at least $\Delta T\geq50$.

To identify phase mismatch between two oscillators, the quantity $\sin{\frac{\varphi_j-\varphi_k}{2}}$ (the sine of the phase semi-difference) is computed.

\begin{itemize}
\item For $\sin{\frac{\varphi_j-\varphi_k}{2}}=1$ anti-phase synchronization is observed ($\varphi_j-\varphi_k = \pi$, i.e., the $j$-th oscillator leads the $k$-th oscillator by $\pi$). We will use the symbol <<{\footnotesize I}>> to denote the anti-phase synchronization: $\varphi_k${\footnotesize I}$\varphi_j$.  
\item $\sin{\frac{\varphi_j-\varphi_k}{2}}=-1$ also corresponds to anti-phase synchronization, however $\varphi_j-\varphi_k = -\pi$, i.e. $k$-th oscillator leads the $j$-th oscillator by $\pi$. $\varphi_j${\footnotesize I}$\varphi_k$.
\item For $\sin{\frac{\varphi_j-\varphi_k}{2}}=0$ in-phase synchronization of two oscillators  is observed.
\item If $\sin{\frac{\varphi_j-\varphi_k}{2}} = \sin{\frac{\varphi_n-\varphi_m}{2}}$ over a certain time interval, then the phase differences for the oscillator pairs $j,k$ and $n,m$ are equal $\varphi_j - \varphi_k = \varphi_n - \varphi_m$ and $(\varphi_j-\varphi_k)-(\varphi_n - \varphi_m)= \text{const}$.
\end{itemize}

Two oscillators are in a frequency synchronized state if their\textit{ mean frequencies} are equal, 
\begin{equation} \label{eq3.21}
\Omega_j = \langle\omega_j\rangle = \Omega_k = \langle\omega_k\rangle,
\end{equation}
where $\omega_j=\dot{\varphi_j}$ and $\omega_k=\dot{\varphi_k}$ are the instantaneous frequencies of the $j$-th and $k$-th oscillators, respectively. The frequencies are defined as

\begin{equation}\label{eq3.2}
\Omega_j=\displaystyle\lim_{T\to\infty}\frac{\varphi_j (T)-\varphi_j (T_0)}{T-T_0},
\end{equation}
where $T_0$ is the transition time.

In addition to synchronization states characterized by the equality of mean frequencies, there exist states of temporary synchronization (over a certain time interval) where \textit{instantaneous frequencies} coincide, 
\begin{equation} \label{eq3.22}
\omega_j = \omega_k.
\end{equation}
Computational experiments have shown that these intervals of coinciding instantaneous frequencies can be quite prolonged. Furthermore, only two instantaneous frequencies can coincide at any given time.

In the course of the study, for various values of the parameter $d$ and randomly selected sets of natural frequencies $\gamma_j$ and initial conditions $\varphi_j(t=0)$ (2000 experiments in total), all possible mean frequency synchronization states for $N=4$ were found. We will separate the sizes of oscillator groups with identical mean frequency by a colon. Along with the fully desynchronized 1:1:1:1 state, four states that are synchronous in mean frequency are possible.

\begin{itemize}
    \item 4:0 -- corresponds to the case, in which all oscillators rotating with the same frequency ($\Omega_1=\Omega_2=\Omega_3=\Omega_4=\Omega^s$).
    \item 3:1 -- one oscillator has a frequency that differs from the others. For example, $\Omega_1=\Omega_2=\Omega_4=\Omega_{124}^s\neq\Omega_3$. This particular realization occurs due to the closeness of the natural frequencies $\gamma_1$, $\gamma_2$, and $\gamma_4$; for a different choice of individual frequencies, oscillators with other indices can synchronize.
    \item 2:2 -- pairwise synchronization. For example, $\Omega_1=\Omega_2 =\Omega_{12}^s\neq\Omega_3=\Omega_4=\Omega_{34}^s$. This situation becomes possible due to the pairwise closeness of the natural frequencies $\gamma_1$ and $\gamma_2$, as well as $\gamma_3$ and $\gamma_4$.
    \item 2:1:1 -- two oscillators are in a synchronized state, while the other two have frequencies that differ from those of the synchronized oscillators and from each other. For example, $\Omega_2=\Omega_4=\Omega_{24}^s\neq\Omega_1\neq\Omega_3$. The reason for the formation of such mismatch is the closeness of the natural frequencies $\gamma_2$ and $\gamma_4$.
\end{itemize}

The 4:0 state represents global synchronization, while the 2:2, 3:1, and 2:1:1 states correspond to cluster synchronization. In the case of global frequency synchronization, two distinct states exist, differentiated by their phase mismatch: a splay state, hereafter denoted as $\text{(4:0)}_1$ (obtained for identical oscillators), characterized by the phase mismatch $|\varphi_2-\varphi_1|=|\varphi_3-\varphi_2|= |\varphi_4-\varphi_3|=|\varphi_1-\varphi_4|=c_1$, where $c_1=\pi/2$; and an in-phase--anti-phase state, hereafter denoted as $\text{(4:0)}_2$, for which, for example, $|\varphi_1-\varphi_4|<c_2$, $|\varphi_2-\varphi_3| <c_2$, $|\varphi_{2}-\varphi_{1} - \pi| <c_2$, $|\varphi_{3}-\varphi_{4} - \pi| <c_2$, where $c_2 \ll 1$. In this state, the first and fourth, as well as the second and third oscillators, are nearly in-phase, while the pair comprising the first and fourth oscillators is in anti-phase with the pair comprising the second and third oscillators.

\begin{figure}
\centering
\includegraphics[width=0.5\textwidth]{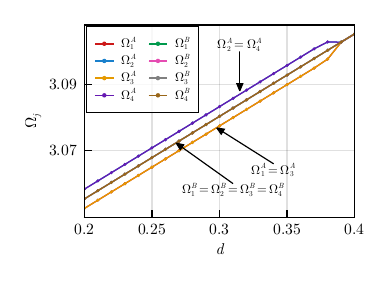}
\caption{Dependence of mean frequencies on the coupling strength. Natural frequencies are: $\gamma_1=3.00268, \gamma_2=3.00821, \gamma_3=3.00233, \gamma_4=3.00847$. The superscripts $A$ and $B$ denote two experiments with different initial conditions. In experiment $A$ 2:2 synchronization state is observed. In experiment $B$ $\text{(4:0)}_2$ synchronization state exists. In the case of 2:2 synchronization, a random choice of natural frequencies leads to the synchronization of the first and third oscillators, and of the second and fourth oscillators.}
\label{fig:Fig_1}
\end{figure}

Fig.~\ref{fig:Fig_1} shows examples of the dependence of the mean frequencies $\Omega_j$ on the coupling parameter $d$. Numerical experiments demonstrated the effect of bistability: for two different sets of initial conditions and a fixed distribution of natural frequencies, the 2:2 and $\text{(4:0)}_2$ synchronization states coexist.  

In the case where the phase differences are constant in time $\varphi_j(t)-\varphi_k(t)=\text{const}$, all mean frequencies are equal, and the formula for calculating the synchronization frequency can be obtained from Eq.~\eqref{eq2.1}. From the equality of phase differences it follows that $\dot{\varphi}_j =\omega_j =\dot{\varphi}_k = \omega_k = \Omega^s$; for all $j,k$, where $\Omega^s$ is the mean synchronization  frequency. Substituting these conditions into Eq.~\eqref{eq2.1}, summing all equations and separating pairs $j,k$ and $k,j$, we obtain the formula:

\begin{equation}\label{eq3.3}
\Omega^s=\displaystyle\frac{1}{N}\sum_{j=1}^N{\gamma_j} - \frac{2d\sin{\alpha}}{N^2}\sum^N_{j<k}{\cos{(\varphi_j-\varphi_k)}}.
\end{equation}

Eq.~\eqref{eq3.3} in the case of nonidentical natural frequencies is valid only for the 4:0 state. This formula also reflects the dependence of the mean frequencies $\Omega_j=\Omega^s$ on the coupling strength $d$, where the term $-\displaystyle\frac{2\sin{\alpha}}{N^2}\sum^N_{j<k}{\cos{(\varphi_j-\varphi_k)}}$ corresponds to the slope of the straight line, and $\displaystyle\frac{1}{N}\sum_{j=1}^N{\gamma_j}$ represents the offset relative to the origin.

A special case of Eq.~\eqref{eq3.3} is the uniform phase distribution (the in-phase--anti-phase synchronization state or splay state). Then the expression for calculating the mean frequency takes the form:

\begin{equation}\label{eq3.4}
\Omega^s=\displaystyle\frac{1}{N}\sum_{j=1}^N{\gamma_j} +\frac{4d\sin{\alpha}}{N^2}
\end{equation}
or, for the number of oscillators under consideration ($N=4$):
\begin{equation}\label{eq3.4(1)}
\Omega^s=\displaystyle\frac{1}{N}\sum_{j=1}^N{\gamma_j} +\frac{d\sin{\alpha}}{4}.
\end{equation}  

Numerical experiments have confirmed that for the 2:2 state (an example is shown in Fig.~\ref{fig:Fig_2}(c)), the pairwise synchronization frequencies with increasing coupling strength can be described by the following expressions:

\begin{equation}\label{eq3.5}
\Omega_{13}^s=\displaystyle\frac{\gamma_1+\gamma_3}{2}+\frac{d\sin{\alpha}}{4},\quad \Omega_{24}^s=\frac{\gamma_2+\gamma_4}{2} + \frac{d\sin{\alpha}}{4}.
\end{equation}

In this state, one pair of mutually synchronized oscillators (in the example in Fig.~\ref{fig:Fig_1}, the second and the fourth) rotates with a frequency larger than that of the other pair of mutually synchronized oscillators (the first and the third).

According to Eq.~\eqref{eq3.4(1)} and Eq.~\eqref{eq3.5}, the synchronization frequencies increase proportionally to the value of the coupling parameter $d$: the stronger the coupling, the faster the oscillators rotate.

\section{Stationary synchronization  state}

We will call a synchronization state \textit{stationary} if it exhibits equality of instantaneous frequencies either always (the 4:0 state) or almost always, except for small time intervals during which the system experiences phase rearrangements (the 2:2 state). In the case of the 2:2 state, after a short desynchronization of instantaneous frequencies, the system returns to its original synchronous state.

\subsection{4:0 state}\label{subsec:4.1}

\begin{figure}
\centering
\includegraphics{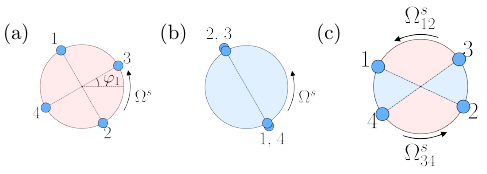}
\caption{Phase representation of the 4:0 and 2:2 states for large values of the coupling parameter $d$. In the case of the 4:0 states ((a) -- splay state $\text{(4:0)}_1$, (b) -- in-phase--anti-phase synchronization state $\text{(4:0)}_2$), all oscillators rotate with constant frequency $\Omega^s$. In the case of the 2:2 state (c), being in anti-phase, the first and second oscillators rotate with frequency $\Omega_{12}^s$, while the third and fourth oscillators, also being in anti-phase, rotate with frequency $\Omega_{34}^s$ ($\Omega_{12}^s > \Omega_{34}^s$).} 
\label{fig:Fig_2}
\end{figure}

In the case of the 4:0 frequency synchronization state, all oscillators rotate with the same constant frequency. Both the means and instantaneous frequencies coincide. Special cases of the 4:0 state include the splay state $\text{(4:0)}_1$ (Fig.~\ref{fig:Fig_2}(a)) and the in-phase--anti-phase synchronization state $\text{(4:0)}_2$ (Fig.~\ref{fig:Fig_2}(b)). Fig.~\ref{fig:Fig_2} illustrates the phase representation of the $\text{(4:0)}_1$, $\text{(4:0)}_2$, for large values of the coupling parameter $d$, where the phase mismatch characteristic of both states are satisfied. Identical coloring of the sectors indicates their equality.

\subsection{2:2 state}\label{subsec:4.2} 

In the case of the 2:2 state, the following pairwise equality of mean frequencies occurs: $\Omega_1=\Omega_2\neq\Omega_3=\Omega_4$.

\begin{figure}
\includegraphics[width=1.0\linewidth]{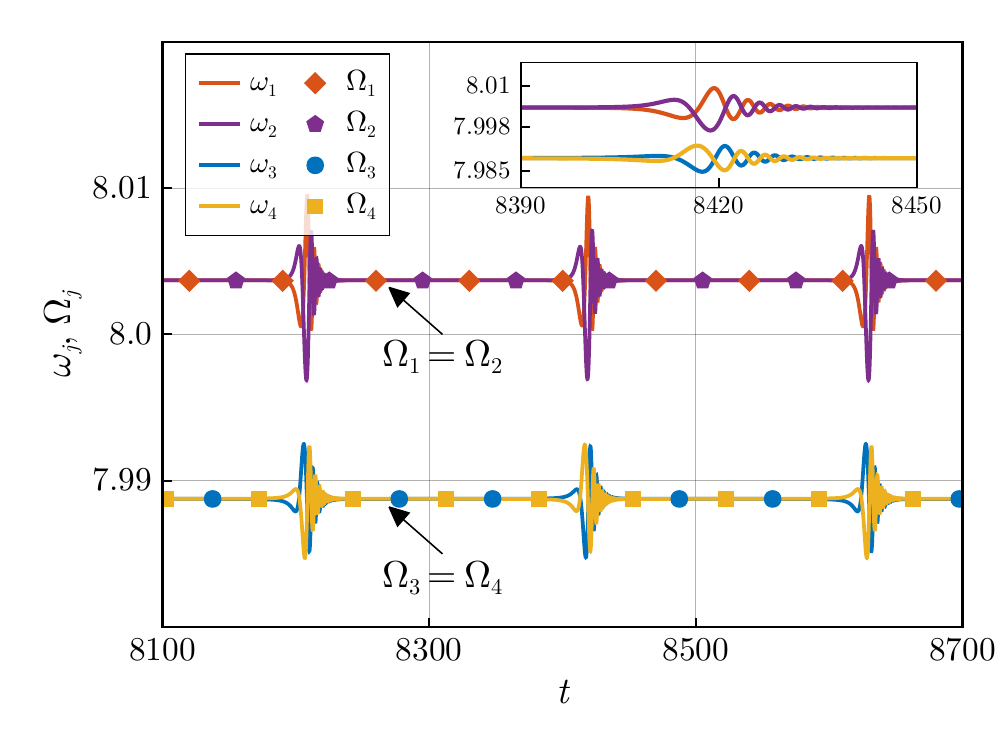}
\caption{\label{fig:Fig_3}Evolution of the mean $\Omega_j$ and instantaneous $\omega_j$ frequencies for the 2:2 synchronization state. The mean frequencies, indicated by markers, were calculated over the interval $t\in[7000,10000]$. Parameters: natural frequencies $\gamma_1=3.00322, \gamma_2=3.00842, \gamma_3=2.99086, \gamma_4=2.99092$, coupling parameter $d=20$. The inset shows the interval of the instantaneous frequencies change. For a detailed description, see the main text.}
\end{figure}

Fig.~\ref{fig:Fig_3} shows the evolution of the means $\Omega_j$ and instantaneous $\omega_j$ frequencies of the 2:2 synchronization state. In this cluster state, one pair of oscillators (the first and second) rotates with a mean frequency larger than the rotation frequency of the other pair of oscillators (the third and fourth).

In this case, throughout the entire plot, except for intervals corresponding to abrupt changes in instantaneous frequencies (for example, the interval $\overline{t}\in(8390,8450)$, see the inset in Fig.~\ref{fig:Fig_3}), the frequency mismatch $\omega_1=\omega_2>\omega_3=\omega_4$ is observed. For each pair of frequency synchronized oscillators, anti-phase synchronization occurs, $\varphi_2-\varphi_1\cong \pi, \varphi_4-\varphi_3\cong\pi$ (Fig.~\ref{fig:Fig_2}(c)). During the interval $\overline{t}$, short desynchronization (the 1:1:1:1 state) occurs due to the disruption of phase mismatch.

\begin{figure}
    \centering
    \includegraphics[width=1.0\linewidth]{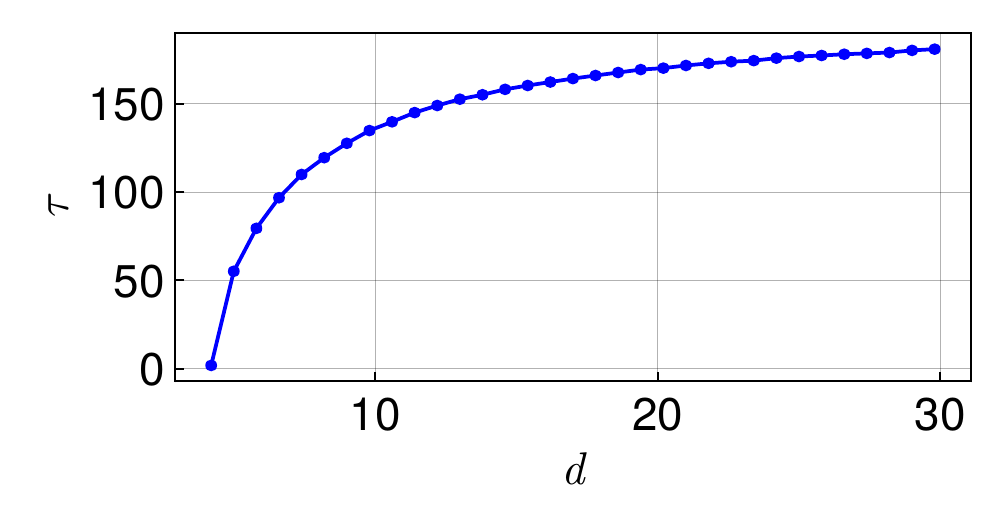}
   \caption{\label{fig:Fig_3_2}Dependence of the "laminar" interval length $\tau$ on the coupling strength $d$ for 2:2 state. Parameters: natural frequencies $\gamma_1=3.00322, \gamma_2=3.00842, \gamma_3=2.99086, \gamma_4=2.99092$. }
\end{figure}

Fig. \ref{fig:Fig_3_2} demonstrates the dependence of the "laminar" interval length $\tau$ on the coupling strength d for the 2:2 state. The curve exhibits a rapid growth at small and moderate $d$ (roughly $d<15$), where an increase in coupling significantly prolongs the intervals of frequency locking, and a slower growth at large $d$ ($d>20$), where $\tau$ approaches values of about 180 within the studied range of $d$.The slowing down of the growth reflects the balance between the synchronizing effect of the coupling and the destabilizing effect of the frequency mismatch: once the coupling is strong enough to keep the oscillators locked, further strengthening does not change the time scale of the slow-fast dynamics. Such dynamics was observed in other synchronous states.

\section{Switching synchronization state}

We will define state as a \textit{switching} state if it exhibits metastable dynamics: a relatively long interval of frequency locking between certain oscillators is observed, followed by a short in time desynchronization, after which the system goes to a new synchronous state -- frequency locking of other pairs of oscillators -- and so on, with the process repeating.

\subsection{3:1 state}\label{subsec:5.1}

In the case of 3:1 cluster synchronization, the following coincidence of mean frequencies occurs: $\Omega_1=\Omega_2=\Omega_3\neq\Omega_4$.

\begin{figure}
\centering
\includegraphics[width=1.0\linewidth]{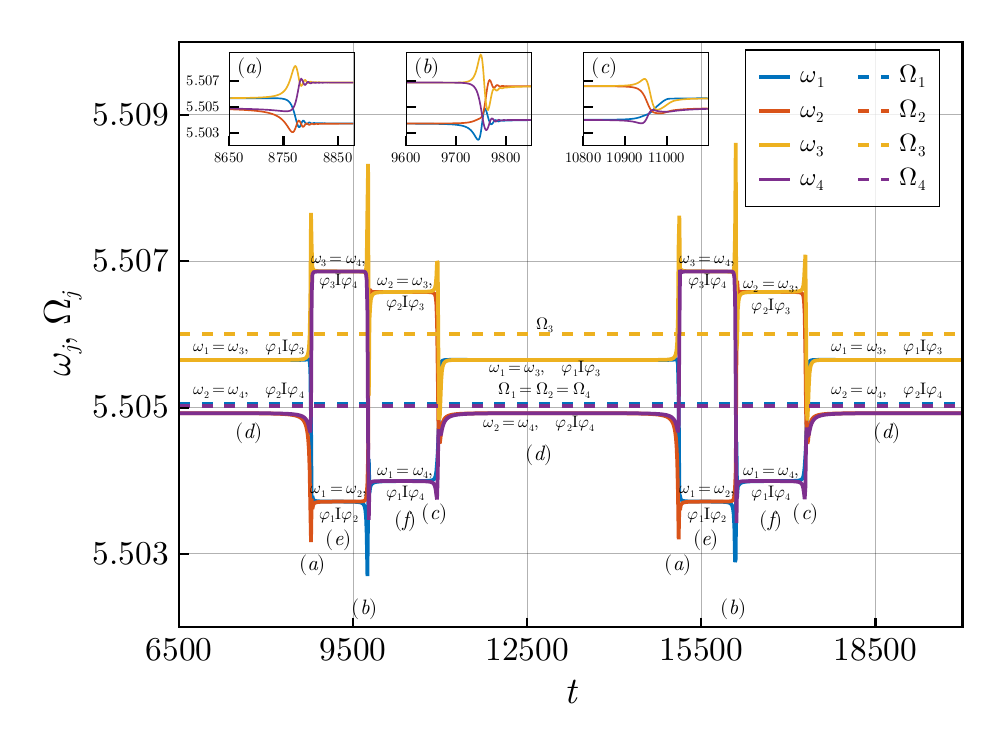}
\caption{Evolution of the mean $\Omega_j$ and instantaneous $\omega_j$ frequencies for the 3:1 synchronization state.  The mean frequencies, indicated by dashed lines, were calculated over the interval $t\in[6800,20000]$. Parameters: natural frequencies $\gamma_1=3.00385, \gamma_2 =3.00571, \gamma_3=3.00958, \gamma_4=3.00627$; coupling strength $d=10$. The insets (a), (b) and (c) show magnified views of the corresponding time intervals where the mismatch between the instantaneous frequencies change. For a detailed description, see the main text.}
\label{fig:Fig_4}
\end{figure}

Fig.~\ref{fig:Fig_4} shows the evolution of the mean frequencies $\Omega_j$ and the instantaneous frequencies $\omega_j=\dot{\varphi}_j$ for a 3:1 synchronization state. In this particular experiment with the chosen intrinsic frequencies $\gamma_j$, the observed state was characterized by $\Omega_1=\Omega_2=\Omega_4\neq\Omega_3$. 

Let us consider the time interval $t \in (7000; 12000)$ and divide it into six subintervals (a)-(f).

In the interval (d) (prior to the first frequency switching), the following conditions: $\omega_1=\omega_3$, $\omega_2=\omega_4$, $\varphi_1${\footnotesize I}$\varphi_3$, $\varphi_2${\footnotesize I}$\varphi_4$. A similar behavior is exhibited by the frequencies and phases in the interval (e) -- $\omega_3=\omega_4$, $\omega_1=\omega_2$, $\varphi_3${\footnotesize I}$\varphi_4$, $\varphi_1${\footnotesize I}$\varphi_2$ and (f) -- $\omega_2=\omega_3$, $\varphi_2${\footnotesize I}$\varphi_3$, $\omega_1=\omega_4$, $\varphi_1${\footnotesize I}$\varphi_4$.

In the intervals (a), (b), and (c), frequency desynchronization occurs, after which the frequencies exhibit distinct behaviors. In the interval (a), the first oscillator initially locks with the fourth, and then with the second. As a result of this frequency switching, $\omega_3=\omega_4>\omega_1=\omega_2$. In the interval (b), the fourth and the second oscillators lock with the first and the third, respectively ($\omega_2=\omega_3>\omega_1=\omega_4$). In the interval (c), the first oscillator locks with the third, and the second with the fourth.

\begin{figure}
\centering
\includegraphics[width=1.0\linewidth]{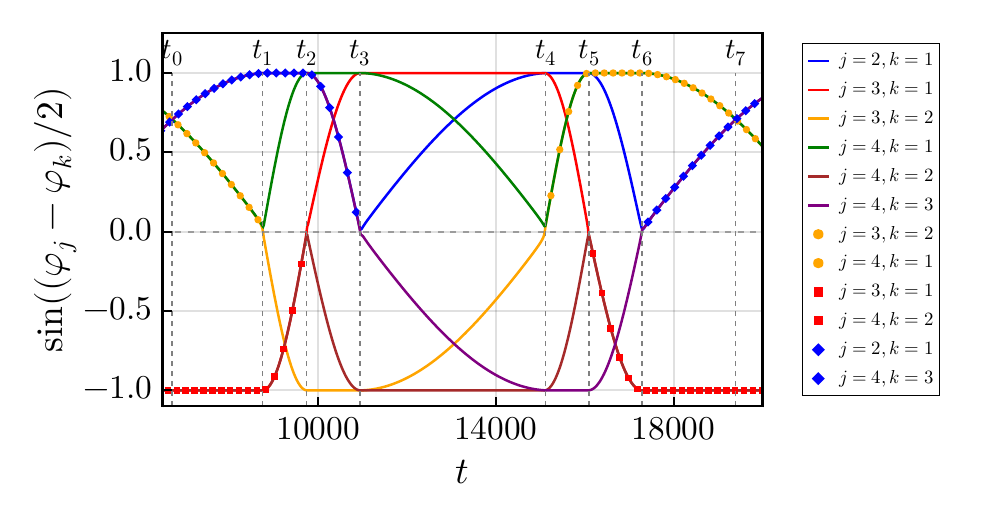}
\caption{Evolution of the sines of the phase semi-differences   $\sin{\frac{\varphi_j-\varphi_k}{2}}$, for the 3:1 state. Parameters are the same as those in Fig.~\ref{fig:Fig_4}. The lines marked with symbols correspond to the equality of the phase differences for two pairs of oscillators. For a detailed description, see the main text.}
\label{fig:Fig_5}
\end{figure}

Fig.~\ref{fig:Fig_5} shows the evolution of the sines of the phase semi-differences for the 3:1 state.

It should be noted that the plot in Fig.~\ref{fig:Fig_5} is periodic with a period $T$, which coincides with the period of the plot of the mean and instantaneous frequencies in Fig.~\ref{fig:Fig_4}.The value of the period is $T=t_4-t_1=6365$.
We divide the plot into seven subintervals: $t_{01}\in(t_0,t_1), t_{12}\in(t_1,t_2), t_{23}\in(t_2,t_3), t_{34}\in(t_3,t_4), t_{45}\in(t_4,t_5), t_{56}\in(t_5,t_6), t_{67}\in(t_6,t_7)$.

In the interval $t_{01}\in(t_0,t_1)$, where $t_0\cong 6720,t_1\cong 8750$, the following equalities: $\sin{\frac{\varphi_2-\varphi_1}{2}}=\sin{\frac{\varphi_4-\varphi_3}{2}}, \sin{\frac{\varphi_4-\varphi_1}{2}}=\sin{\frac{\varphi_3-\varphi_2}{2}}, \sin{\frac{\varphi_3-\varphi_1}{2}}=\sin{\frac{\varphi_4-\varphi_2}{2}}=-1$. The last equality means that the oscillators form anti-phase pairs: the third with the first, and the fourth with the second, with $\varphi_1>\varphi_3,\varphi_2>\varphi_4$. Fig.~\ref{fig:Fig_5} shows the plot of the sine of the phase semi-difference $\sin{\frac{\varphi_j-\varphi_k}{2}}$, whose indices are listed in reverse lexicographical order. When describing the anti-phase states, we will list first in the pair the oscillator with the larger phase ($\varphi_1${\footnotesize I}$\varphi_3,\varphi_2${\footnotesize I}$\varphi_4$, in this case). Then, in the case of lexicographical order, the difference is negative, otherwise it is positive. Since, in this time interval, the rotation frequencies of the first and the third oscillators are larger than those of the second and the fourth, one pair overtakes the other; the fourth and the third, and the second and the first oscillators gradually converge, while the fourth and the first, and the third and the second, conversely, diverge. This leads to in-phase--anti-phase synchronization ($\sin{\frac{\varphi_4-\varphi_1}{2}}=\sin{\frac{\varphi_3-\varphi_2}{2}}=0$) at time $t_1$ (in Fig.~\ref{fig:Fig_4}, moment $t_1$ belongs to the interval (a)), after which a frequency switch occurs -- a change in the mismatch of the instantaneous frequencies.

In the interval $t_{12}\in(t_1,t_2)$, where $t_1\cong 8750,t_2\cong 9740$, the third and the fourth oscillators rotate faster than the first and the second ($\omega_3=\omega_4>\omega_1=\omega_2$). Also, for these pairs $\sin{\frac{\varphi_2-\varphi_1}{2}}=\sin{\frac{\varphi_4-\varphi_3}{2}}=1, \sin{\frac{\varphi_3-\varphi_1}{2}}=\sin{\frac{\varphi_4-\varphi_2}{2}},\varphi_2${\footnotesize I}$\varphi_1,\varphi_4${\footnotesize I}$\varphi_3$. At time $t_2$ (in Fig.~\ref{fig:Fig_4}, moment $t_2$ belongs to the interval (b)), in-phase--anti-phase synchronization ($\sin{\frac{\varphi_3-\varphi_1}{2}}=\sin{\frac{\varphi_4-\varphi_2}{2}}=0$) is achieved, accompanied by a frequency switch.

In the interval $t_{23}\in(t_2,t_3)$, where $t_2\cong 9740$, $t_3\cong 10940$, the second and the third oscillators have larger frequency than the first and the fourth ($\omega_2=\omega_3>\omega_1=\omega_4$) with $\varphi_4$ {\footnotesize I}$\varphi_1,\varphi_2${\footnotesize I}$\varphi_3$. As in previous intervals, at time $t_3$ (in Fig.~\ref{fig:Fig_4}, the moment $t_3$ belongs to the interval (c)), in-phase--anti-phase synchronization is achieved ($\sin{\frac{\varphi_2-\varphi_1}{2}}=\sin{\frac{\varphi_4-\varphi_3}{2}}=0$), accompanied by another frequencies switch.

In the interval $t_{34}\in(t_3,t_4)$, where $t_3\cong 10940$, $t_4\cong 15115$, similarly to the interval $t_{01}$, the third and the first oscillators have larger frequency than the second and the fourth ($\omega_1=\omega_3>\omega_2=\omega_4$), with the third and the first, as well as the fourth and the second oscillators being in anti-phase. However, $\sin{\frac{\varphi_3-\varphi_1}{2}}=1$, $\sin{\frac{\varphi_4-\varphi_2}{2}}=-1$, or $\varphi_3${\footnotesize I}$\varphi_1,\varphi_2${\footnotesize I}$\varphi_4$. That is, due to successive frequency switches in pairs of oscillators with larger and lesser frequencies, at time $t_3$ the third oscillator becomes anti-phase with the first, and now it is the one leading in phase $\varphi_3>\varphi_1$).

A similar switch is presented in the interval $t_{45}\in(t_4,t_5)$, where $t_4\cong 15115$, $t_5\cong 16100$. Relative to the interval $t_{12}$, where $\sin{\frac{\varphi_2-\varphi_1}{2}}=\sin{\frac{\varphi_4-\varphi_3}{2}}=1$, the equality condition splits into two $\sin{\frac{\varphi_2-\varphi_1}{2}}=1$, $\sin{\frac{\varphi_4-\varphi_3}{2}}=-1$ ($\varphi_2-\varphi_1=\pi, \varphi_3-\varphi_4=\pi$). Also, in the interval $t_{45}$, equality of distances between the fourth and the first, the third and the second oscillators is observed ($\sin{\frac{\varphi_4-\varphi_1}{2}}=\sin{\frac{\varphi_3-\varphi_2}{2}}$).

The evolution of the sines of the phase semi-differences in the interval $t_{56}\in(t_5,t_6)$, where $t_5\cong 16100$, $t_6\cong 17290$, is comparable to the behavior in the interval $t_{23}$. However, if in the interval $t_{23}$ the equality splits into two, then in this case, conversely, a merging occurs. In the interval $t_{23}$, the equalities hold: $\sin{\frac{\varphi_4-\varphi_1}{2}}=1$, $\sin{\frac{\varphi_3-\varphi_2}{2}}=-1$ ($\varphi_4-\varphi_1=\pi,\varphi_2-\varphi_3=\pi$); in the interval $t_{56}$ -- $\sin{\frac{\varphi_4-\varphi_`}{2}}=\sin{\frac{\varphi_3-\varphi_2}{2}}=1$ $(\varphi_4-\varphi_1=\varphi_3-\varphi_2=\pi$), that is, the phase difference between the third and the second oscillators changed from $-\pi$ to $\pi$. Also $\sin{\frac{\varphi_3-\varphi_1}{2}}=\sin{\frac{\varphi_4-\varphi_2}{2}}$.

In the interval $t_{67}\in(t_6,t_7)$, where $t_6\cong 17290$, $t_7\cong 19390$, the following equalities hold: $\sin{\frac{\varphi_2-\varphi_1}{2}}=\sin{\frac{\varphi_4-\varphi_3}{2}}$, $\sin{\frac{\varphi_4-\varphi_1}{2}}=\sin{\frac{\varphi_3-\varphi_2}{2}}$, $\sin{\frac{\varphi_3-\varphi_1}{2}}=\sin{\frac{\varphi_4-\varphi_2}{2}}$, and the oscillators are in anti-phase $\varphi_1${\footnotesize I}$\varphi_3,\varphi_2${\footnotesize I}$\varphi_4$. The behavior of the sine of the phase semi-differences is the same as behavior in the interval $t_{01}$.

It should be noted that, despite the 3:1 mean frequency synchronization state, the plot in Fig.~\ref{fig:Fig_5} (excluding short desynchronization intervals) exhibits an alternation of 2:2 instantaneous synchronization states.

It should be noted that there are time intervals in which the average synchronization frequency of several oscillators is higher or lower than the same instantaneous frequencies. For example, see Fig.~\ref{fig:Fig_4}. Analogous behavior has been observed in other cases (2:1:1 frequency synchronization).

\subsection{2:1:1 state}

In the case of the 2:1:1 synchronization state, the following coincidence of mean frequencies is observed: $\Omega_1=\Omega_2\neq\Omega_3\neq\Omega_4$.

\begin{figure}
\centering
\includegraphics[width=1.0\linewidth]{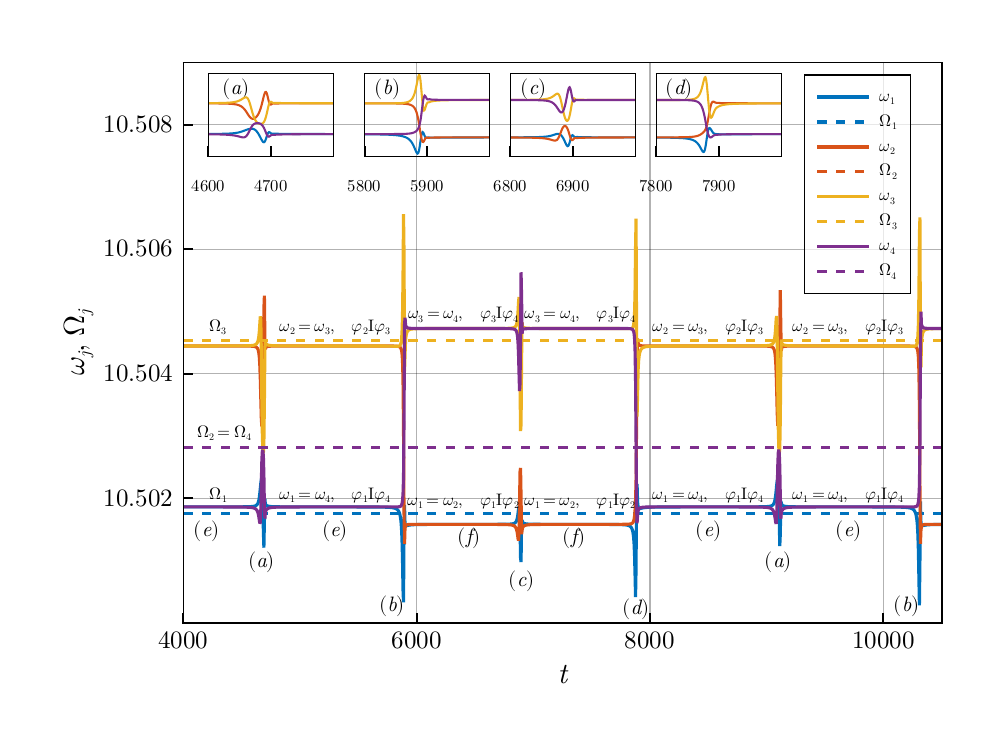}
\caption{Evolution of the mean $\Omega_j$ and instantaneous $\omega_j$ frequencies for the 2:1:1 state. The mean frequencies, indicated by dashed lines, were calculated over the interval $t\in[4000, 10500]$. Parameters: natural frequencies $\gamma_1=3.00385$, $\gamma_2=3.00571$, $\gamma_3=3.00958$, $\gamma_4=3.00627$; coupling strength $d=30$. The insets (a), (b), (c) and (d) show magnified views of the corresponding time intervals where the mismatch between the instantaneous frequencies change. For a detailed description, see the main text.}
\label{fig:Fig_6}
\end{figure}

Let us consider the evolution of the means $\Omega_j$ and instantaneous $\omega_j$ frequencies for the 2:1:1 state (Fig.~\ref{fig:Fig_6}).

The behavior of the oscillators in this case is the same as their behavior in the 3:1 state. In the interval (e), the following mismatch: $\omega_2=\omega_3>\omega_1=\omega_4$, $\varphi_2${\footnotesize I}$\varphi_3$, $\varphi_1${\footnotesize I}$\varphi_4$. In the interval (f), the conditions are: $\omega_3=\omega_4>\omega_1=\omega_2$, $\varphi_2${\footnotesize I}$\varphi_3$, $\varphi_1${\footnotesize I}$\varphi_4$. The intervals (a), (b), (c), (d) are characterized by the presence of desynchronization; however, in this case, it is not always accompanied by a switch of the frequency synchronization states. For example, during the intervals (a) and (c), the relationships $\omega_1=\omega_4>\omega_2=\omega_3$ and $\omega_3=\omega_4>\omega_1=\omega_2$ are maintained for each interval, respectively. In contrast, the intervals (b) and (d) exhibit a switch of the frequency synchronization states. In the interval (b), the first and the second, and the third and the fourth oscillators are synchronized: $\omega_1=\omega_2$, $\omega_3=\omega_4$, with $\omega_3=\omega_4>\omega_1=\omega_2$. The interval (d) is characterized by a switch of synchronization states opposite to that in (b) -- the second and the third, and the first and the fourth oscillators are synchronized: $\omega_2=\omega_3$, $\omega_1=\omega_4$, with $\omega_2=\omega_3>\omega_1=\omega_4$.

\begin{figure}
\centering
\includegraphics[width=1.0\linewidth]{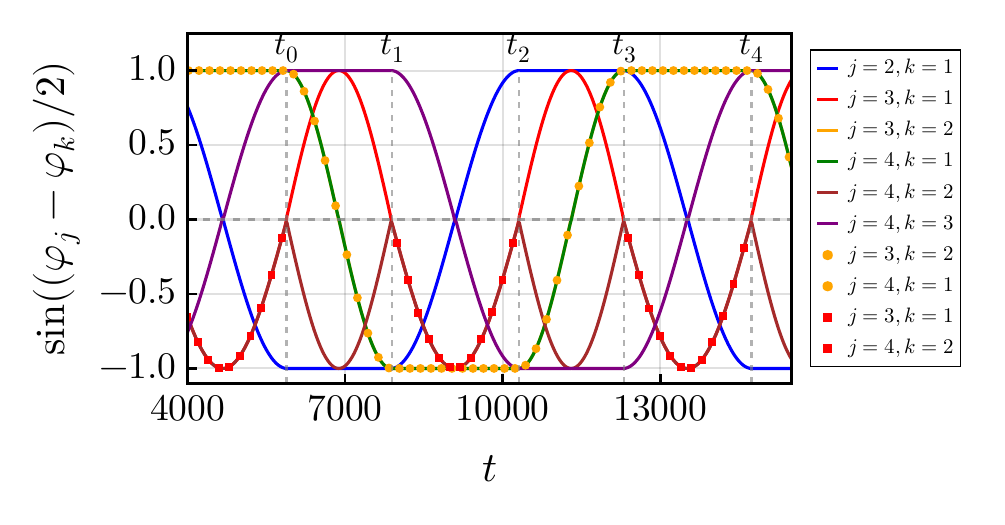}
\caption{Evolution of the sines of the phase semi-difference   $\sin{\frac{\varphi_j-\varphi_k}{2}}$ for the 2:1:1 state. Parameters are the same as in Fig.~\ref{fig:Fig_6}. Lines marked with markers correspond to the equality of phase differences for two oscillator pairs. For a detailed description, see the main text.}
\label{fig:Fig_7}
\end{figure}

Fig.~\ref{fig:Fig_7} shows the plot of the evolution of the sines of phase semi-differences. Similar to the 3:1 state, the evolution is periodic, with a period $T = t_3 - t_1 = 4420$.

Consider the time interval $t_{04}\in(t_0,t_4)$ and divide it into four subintervals: $t_{01}\in(t_0,t_1),t_{12}\in(t_1,t_2),t_{23}\in(t_2,t_3),t_{34}\in(t_3,t_4)$.

In the interval $t_{01} \in (t_0, t_1)$, where $t_0 \cong 5885$, $t_1 \cong 7890$, the oscillators rotate in pairs, with the third and the fourth faster than the first and the second ($\omega_3 = \omega_4 > \omega_1 = \omega_2$). The following equalities also hold: $\sin{\frac{\varphi_4-\varphi_3}{2}}=1$, $\sin{\frac{\varphi_2-\varphi_1}{2}} = -1$, i.e., $\varphi_4${\footnotesize I}$\varphi_3$, $\varphi_2${\footnotesize I}$\varphi_1$. Throughout the interval, the same equality $\sin{\frac{\varphi_4-\varphi_1}{2}} = \sin{\frac{\varphi_3-\varphi_2}{2}}$ is observed between the first and the fourth, and the third and the second oscillators. Here, $\sin{\frac{\varphi_4-\varphi_1}{2}}$, $\sin{\frac{\varphi_3-\varphi_2}{2}}$ gradually decrease until reaching a splay state, and then further until in-phase--anti-phase synchronization is achieved. The passing of $\sin{\frac{\varphi_j-\varphi_k}{2}}$ through zero indicates that one oscillator leads in phase to the other (in this case, the third leads the second and the fourth leads the first). Thus, one pair overtakes the other, exhibiting anti-phase synchronization or, at certain times, a splay state where $\sin{\frac{\varphi_3-\varphi_2}{2}} = \sin{\frac{\varphi_4-\varphi_1}{2}} = \sin{\frac{\varphi_4-\varphi_2}{2}}$, or in-phase--anti-phase synchronization (at time $t_1$), where $\sin{\frac{\varphi_3-\varphi_1}{2}} = \sin{\frac{\varphi_4-\varphi_2}{2}} = 0$. The instantaneous frequency switch intervals in Fig.~\ref{fig:Fig_6} coincide with the phase mismatch switch intervals in Fig.~\ref{fig:Fig_7}. Unlike the 3:1 state, in this case, the frequency synchronization state switches occur not after every first attainment of in-phase--anti-phase synchronization, but after every second one.

In the interval $t_{12} \in (t_1, t_2)$, where $t_1 \cong 7890, t_2 \cong 10305$, there also exist two time instances corresponding to the splay state and one instance corresponding to the in-phase--anti-phase synchronization state. At all other points of the interval, anti-phase synchronization is observed, satisfying the equalities: $\sin{\frac{\varphi_3-\varphi_1}{2}} = \sin{\frac{\varphi_4-\varphi_2}{2}}, \sin{\frac{\varphi_4-\varphi_1}{2}} = \sin{\frac{\varphi_3-\varphi_2}{2}} = -1$, which means $\varphi_1${\footnotesize I}$\varphi_4 , \varphi_2${\footnotesize I}$\varphi_3$.

In the interval $t_{23} \in (t_2, t_3)$, where $t_2 \cong 10305$, $t_3 \cong 12315$, the evolution of the sines of the phase semi-differences is the same as the behavior in the interval $t_{01}$. The second and the first, and the third and the fourth oscillators are in anti-phase; however, compared to the interval $t_{01}$, the equalities $\sin{\frac{\varphi_2-\varphi_1}{2}}=1$, $\sin{\frac{\varphi_4-\varphi_3}{2}}= -1$ hold, which means $\varphi_2${\footnotesize I}$\varphi_1$, $\varphi_3${\footnotesize I}$\varphi_4$.

In the interval $t_{34} \in (t_3, t_4)$, where $t_3 \cong 12315$, $t_4 \cong 14735$, the following equalities: $\sin{\frac{\varphi_3-\varphi_1}{2}} = \sin{\frac{\varphi_4-\varphi_2}{2}}$, $s_{41} = \sin{\frac{\varphi_3-\varphi_2}{2}}$, corresponding to $\varphi_4${\footnotesize I}$\varphi_1$, $\varphi_3${\footnotesize I}$\varphi_2$. The second and the first oscillators approach each other, as do the third and the first, and the fourth and the second oscillators, tending towards a splay state and then towards in-phase--anti-phase synchronization ($\varphi_4${\footnotesize I}$\varphi_1$, $\varphi_3${\footnotesize I}$\varphi_2$, $\varphi_1${\footnotesize I}$\varphi_3$, $\varphi_2${\footnotesize I}$\varphi_4$). After this, the third and the fourth, the third and the first, and the fourth and the second oscillators begin to separate, while the first and the second oscillators approach each other. This leads to the appearance of in-phase--anti-phase synchronization at time $t_4$ ($\varphi_4${\footnotesize I}$\varphi_1$, $\varphi_3${\footnotesize I}$\varphi_2$, $\varphi_4${\footnotesize I}$\varphi_3$, $\varphi_1${\footnotesize I}$\varphi_2$).

Similarly to the case of the 3:1 mean frequency synchronization state, the 2:1:1 state is characterized by an alternation of instantaneous 2:2 synchronization states.

\section*{Conclusion}
This paper investigates synchronization states in an ensemble of four phase oscillators with repulsive coupling and nonidentical natural frequencies. All possible frequency synchronization states have been identified (4:0~--~where all mean frequencies coincide, corresponding to the global synchronization state; 3:1~--~three mean frequencies coincide; 2:2~--~pairs of mean frequencies coincide; 2:1:1~--~two mean frequencies coincide). These states can be classified into two types: stationary and switching. The stationary type includes the 4:0 and 2:2 states, characterized by constant synchronization instantaneous frequencies. The 4:0 state has two types: the splay state (denoted as $\text{(4:0)}_1$ and obtained for identical oscillators) and the in-phase--anti-phase synchronization state (denoted as $\text{(4:0)}_2$). In the 2:2 state, after short in time desynchronization, the system returns to its original synchronous state, preserving groups with equal instantaneous frequencies. In the case of the 2:1:1 and 3:1 switching synchronization states, an alternation of 2:2 instantaneous states are observed. A novel effect of collective dynamics -- self-organized emergence of slow–fast behavior in the ensemble of elements which, if decoupled, exhibit uniform temporal evolution -- was demonstrated. According to the derived analytical expressions, synchronization frequencies increase proportionally with coupling strength: stronger coupling leads to faster oscillator rotation.

\section*{Acknowledgments}

The work was supported by the Russian Science Foundation (project no. 23-12-00180-P -- section II, III and project no. 22-12-00348-P -- section IV, V).

\section*{Data availability}
The data that support the findings of this study are available from the corresponding author upon reasonable request.

\nocite{*}
\bibliography{aipsamp.bib}

\end{document}